\documentclass{article}

\usepackage{PRIMEarxiv}
\usepackage{caption}
\usepackage{subfigure}
\usepackage[utf8]{inputenc} 
\usepackage[T1]{fontenc}    
\usepackage{hyperref}       
\usepackage{url}            
\usepackage{amsfonts}       
\usepackage{nicefrac}       
\usepackage{microtype}      
\usepackage{fancyhdr}       
\usepackage{graphicx}       
\usepackage[numbers]{natbib}
\usepackage{savesym}
\savesymbol{fax}
\savesymbol{Hermaphrodite}
\usepackage[show]{chato-notes}
\usepackage{color}
\usepackage[at]{easylist}
\usepackage{array}
\usepackage{algorithmic}
\usepackage{algorithm}
\usepackage{enumitem}
\usepackage{csvsimple}
\usepackage{amsmath}
\usepackage{lscape}
\usepackage{multirow}
\usepackage{hhline}
\usepackage{pdfpages}
\usepackage{textcomp}
\usepackage{graphics}
\usepackage{booktabs,caption}
\usepackage[flushleft]{threeparttable}
\usepackage{pifont}
\usepackage[at]{easylist}
\usepackage{amssymb}
\newcolumntype{M}{>{\arraybackslash}m{1.2cm}}
\newcolumntype{O}{>{\arraybackslash}m{0.8cm}}
\newcolumntype{P}{>{\arraybackslash}m{5cm}}
\newcolumntype{Q}{>{\arraybackslash}m{11cm}}
\newcommand{\new}[1]{\textcolor{black}{#1}}
\newcommand{\final}[1]{\textcolor{black}{#1}}
\newcommand{\bof}[1]{\textcolor{black}{#1}}
\newcommand{\cmn}[1]{\textcolor{black}{#1}}

\DeclareMathOperator{\follows}{follows}

\title{Leveraging Users' Social Network Embeddings for Fake News Detection on Twitter}

\author{
  Ting Su, Craig Macdonald, Iadh Ounis \\
  University of Glasgow \\
  Glasgow, UK\\
  \texttt{t.su.2@research.ac.uk, \{craig.macdunald, iadh.ounis\}@glasgow.ac.uk}  \\
}

\begin{document}
\maketitle

\begin{abstract}
Social networks (SNs) are increasingly important sources of news for many people. 
The online connections made by users allows information to spread more easily than traditional news media (e.g., newspaper, television).
However, they also make the spread of fake news easier than in traditional media, especially through the users' social network connections.
In this paper, we focus on investigating if the SNs' users connection structure can aid fake news detection on Twitter. 
In particular, we propose to embed users based on their follower or friendship networks on the Twitter platform, so as to identify the groups that users form. 
Indeed, by applying unsupervised graph embedding methods on the graphs from the Twitter users' social network connections, we observe that users engaged with fake news are more tightly clustered together than users only engaged in factual news.
Thus, we hypothesise that the embedded user's network can help detect fake news effectively. 
Through extensive experiments using a publicly available Twitter dataset, our results show that applying graph embedding methods on SNs, using the user connections as network information, can indeed classify fake news more effectively than most language-based approaches. 
Specifically, we observe a significant improvement over using only the textual information (i.e., TF.IDF or a BERT language model), as well as over models that deploy both advanced textual features (i.e., stance detection) and complex network features (e.g., users network, publishers cross citations). 
We conclude that the Twitter users' friendship and followers network information can significantly outperform language-based approaches, as well as the existing state-of-the-art fake news detection models that use a more sophisticated network structure, in classifying fake news on Twitter.
\end{abstract}


\section{Introduction}
\label{sec:intro}
In recent years, the term ``fake news" has been used abundantly, by both politicians and journalists. 
The existence of large amounts of information that members of the public can't or won't trust, or that can easily confuse and mislead the general public, can cause public distrust in journalism. Such a reality is already unfolding, as the reputation of news outlets and news media has been in doubt amid a global distrust crisis~\cite{latkin2020assessment}. 

\begin{figure}[tb]
    \centering
    \includegraphics[width=0.7 \textwidth]{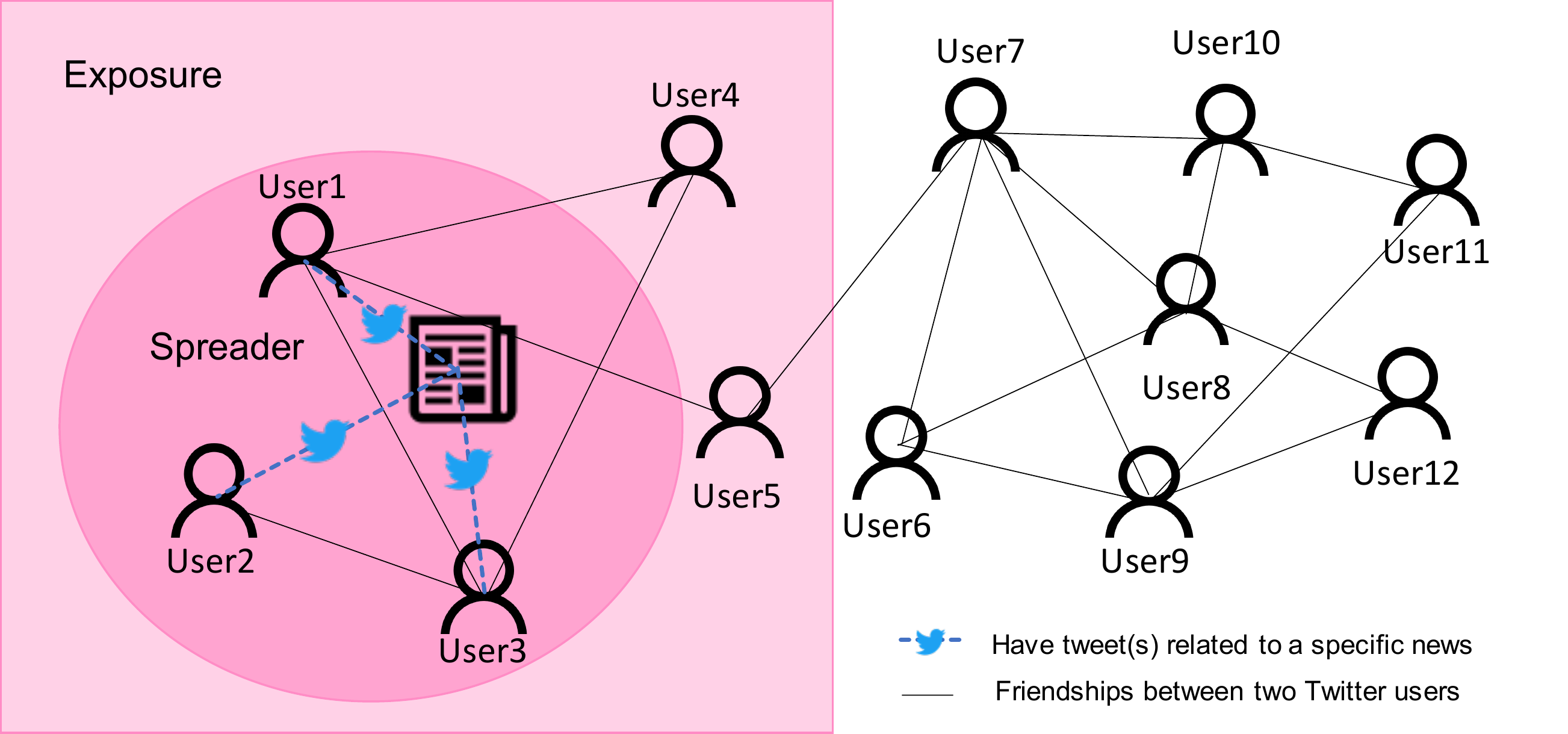}
    \captionsetup{width=0.65\linewidth}
    \caption{An example showing that users are connected on the Twitter platform, while only a subset of users engage with a specific news article.} 
    \label{fig:illu}\vspace{-2\baselineskip}
\end{figure}

Similarly, polarisation in political and scientific debates are observed more frequently~\cite{prior2013media, pierson2020madison}, since fake news~\footnote{In this paper, we use ``news" to describe all types of news format - articles, tweets, claims, etc. 
} can amplify and reinforce false information through repeated exposure~\cite{polage2012making,pennycook2018prior}.

Cho et al.~\cite{cho2020search} showed that one possible reason for the increasing polarisation is that the selected contents delivered by search engines (e.g., Google, YouTube) and social media platforms (e.g., Facebook, Twitter) are individually tailored to the users' specific interests, thus creating \textit{filter bubbles} that reinforce their existing beliefs. 
Such an observation echoes the findings of Ling~\cite{ling2020confirmation}, who showed that individuals' false beliefs are emphasised by repeated exposure and a selective focus because of the \textit{confirmation bias} effect.
Indeed, the confirmation bias states that individuals are more inclined to interact with information that aligns or confirms their existing beliefs, while avoid confrontational information that challenge their existing beliefs.
Furthermore, Yoo~\cite{yoo2007ideological} identified \textit{echo chamber} effect in social media platforms and showed that this facilitates rumours and fake news being created and circulating within small and specific groups, before being spread further. Echo chamber effect in social media can be described as a subset of like-minded users grouping together, with little interaction with dissimilarly-minded users, to preserve their beliefs and avoid confrontation. Figure~\ref{fig:illu} illustrates that a news article might only reach a subset of users, with it being circulated and discussed by like-minded group members. 

\looseness -1 However, Lewandowsky at 
 al.~\cite{lewandowsky2012misinformation} argued that the misinformation crisis should be considered and evaluated as a public concern, especially with the popularity of social media. 
Indeed, increasing efforts~\cite{hamidian2016rumor,kasami1966efficient,feng2012syntactic,ciampaglia2015computational} are made in recent years to develop methods that can help scholars and journalists detect fake news.
These methods generally deploy machine learning methods using text features (i.e., TF.IDF~\cite{ma2016detecting}, word2vec~\cite{jang2019word2vec}), information cascade features~\cite{liang2015rumor,ruchansky2017csi}, metadata extracted from tweets, as well as entity and knowledge base methods~\cite{ciampaglia2015computational,su2019entity}. 

However, the above mentioned methods largely focus on the textual features, statistics of tweets and users engaged in the topic, without paying much attention to the SN structure around users -- indeed, users on social media form their unique society and are largely distinguishable based on their interests and beliefs. One recent related study~\cite{nguyen2020fang} that investigated the user network's role in detecting fake news deployed complex network structures (e.g., citation network, publication network, stances) and achieved significant improvements in accuracy at classifying a news article as fake or not, compared to the existing state-of-the-art models.
Inspired by this, we propose a user network embedding method to represent each Twitter user in a lower dimensional space, based on their connections with other users within the platform, and projecting the social network structure as a linked graph.
We argue that this graph structure can aid in detecting clusters of users engaging in fake news, and can assist the fake news detection task on Twitter. Thus, the main contributions of this paper are as follows:

\begin{itemize}
    \item We propose to incorporate the idea of echo chamber effect into automatic fake news detection task. Specifically, we hypothesise that training user network embeddings without the prior knowledge of users' engagements with fake news can help identifying user communities that frequently engage in and spread fake news, and facilitate the fake news detection in social media.
    
    \item We propose a User Network Embedding Structure (UNES) model, which performs fake news classification on Twitter through the use of graph embeddings to represent Twitter user's social network structure. Compared to the approach of~\cite{nguyen2020fang}, UNES does \textbf{not} require any pre-annotated data (e.g., user type (individual users or publishers), users stance, and if they have engaged with fake news before).
    
    \item We observe that the user embeddings generated by UNES exhibit a clustering effect between users who engage with fake news and users who solely engage with factual news, despite not having knowledge if the users have engaged with fake news before.
    
    \item We also show that using the social network's user connections alone to build network embeddings, and using only users that engaged with the news when representing such news, can significantly outperform the existing state-of-the-art fake news detection approaches that use both textual features and complex social network features.
\end{itemize}

This paper is structured as follows: Section~\ref{sec:related} describes the fake news detection methods proposed in the literature, and highlights the limitations in these approaches, while also providing a background on graph embeddings, particularly in the social networks domain;
Section~\ref{sec:method} formally states the task to be tackled and and describes our proposed UNES model for fake news detection; 
The used experimental setup and obtained results for fake news detection using the UNES model are provided in
Sections~\ref{sec:setup} and \ref{sec:results}, respectively;
Concluding remarks follow in Section~\ref{sec:conclusions}.

\section{Related Work}
\label{sec:related}
There is an extensive body of research on fake news detection in social media. 
In this section, we focus on commonly used fake detection approaches, as well as survey recent advances in graph embedding models. Then, we identify some limitations of the existing methods, and position our study in relation to these methods.

\subsection{Fake News Detection}
\label{sec:background1}
Research in fake news -- sometimes also called rumours, misinformation, or disinformation -- has attracted increasing attention in recent years. 
Identifying fake news is an important step in informing the general public about the widespread of false news. 
Thus, we first focus on existing methods that aim to tackle the task of identifying fake news in social media.

\textbf{Fake news detection using textual information.} 
Text representations are widely used in fake news detection. 
There are several methods used to represent text. 
The most commonly used methods in fake news detection include the bag-of-words (BoW) method (e.g.\ TF.IDF~\cite{sparck1972statistical}), parts-of-speech (POS tagging~\cite{brill1992simple}), and word embedding (e.g.\ Word2Vec~\cite{mikolov2013efficient}) representations. For example, Ma et al.~\cite{ma2016detecting} used a typical BoW representation (TF.IDF) to represent tweets. 
Deeper NNs, such as long short term memory (LSTM) models and bidirectional LSTMs (BiLSTMs), can capture the semantic meaning of the text based on nearby tokens, and is used in fake news detection tasks~\cite{favano2019theearthisflat}. Kochkina et al.~\cite{kochkina2017turing} proposed BranchLSTM to analyse tweet threads (tweets, retweets, comments) sequentially. BranchLSTM yielded more accurate results in identifying fake news, than those without the sequential information of tweet threads.
Moreover, recent research has shown that attention-based neural network language models (e.g.\ ELMO~\cite{peters2018deep}, BERT~\cite{devlin2018bert}) can benefit from both the complex neural network structure and the large pre-training corpus. 
As the state-of-the-art pre-trained language model, the BERT model can be fine-tuned to tailor to the needs of a specific task, and has been shown to consistently outperform other models in many tasks~\cite{lee2020biobert,yilmaz2019applying}, as well as in fake news detection tasks~\cite{su2019ensembles}.

Many scholars also used other handcrafted textual features (e.g.\ Number of @, \#, exclamation marks, first-person pronouns~\cite{qazvinian2011rumor,zhao2015enquiring,hamidian2016rumor}).
Deep syntactical analysis approaches~\cite{kasami1966efficient} are also useful in detecting fake news~\cite{feng2012syntactic}.

\textbf{Network Analysis.}
Networks, usually represented as a linked graph, have data points linking to one or more other data points through information or relationships. 
Knowledge graphs and social networks are the two main types of networks that are particularly rich in conveying additional information that textual features may not capture, when detecting fake news online~\cite{shu2017fake,zhou2019network}.

Knowledge Graphs use a linked structure that consists of entities and their relationships (e.g.\ \textit{$\langle$Beethoven, composer\_of, Moonlight Sonata$\rangle$}, where \textit{Beethoven} and \textit{Moonlight Sonata} are the two entities and \textit{composer\_of} is the relationship between the two entities). 
By extracting entities and relationships between entities from the text, and comparing this information to well-established knowledge graphs, it is possible to estimate the likelihood of the news being factual or fake.
For example, Ciampaglia at a.~\cite{ciampaglia2015computational} found that by calculating the shortest path of the subject and object in a knowledge base, where the subject and the object are from the triplet ($\langle subject,predicate,object\rangle$) extracted from the article, they can reach an accuracy as high as 90\% in classifying fake and non-fake news. 

Social networks (such as Twitter) are platforms where people can connect with each other online, and share information and news with their followers and friends. 
The links between people on Twitter can be explicitly categorised into two types: \textit{followers} or \textit{friends}.
Researchers have developed a range of features that use the rich information that social networks provide, to help identify fake news.
Some of the examples are as follows: simple user features (e.g, \# of followers, \# of friends, verified or not, description)~\cite{liang2015rumor}, relations between users (e.g., in the same region, engaged with same tweet/URL)~\cite{ruchansky2017csi}, relations between tweets (e.g., replies, retweets, likes, viewpoints conflicts)~\cite{jin2016news}.

However, all the above mentioned methods that include users information as one of the features, only consider statistical user information, but ignore the potential network information that may help in identifying the social group a particular user belongs to, and the stance that this group has on the particular issue. 
\new{Recently, researchers started analysing the \bof{user} network structure of social media, such as \bof{the} user network on Twitter \bof{or the} channel network on YouTube.}
For example, R{\"o}chert et al.~\cite{rochert2021networked} studied the user network structure of the YouTube channels. They found that the channels and individuals propagating fake news on YouTube are usually integrated in heterogeneous discussion networks that involve factual content more than misinformation. 
Similarly, Sosnkowsk et al.~\cite{sosnkowski2021analysis} showed that changes in the users' network structure on Twitter can help detect the change of political opinions among users. 
Nguyen et al.~\cite{nguyen2020fang} proposed the FActual News Graph (FANG) model, and showed that network information can be beneficial in detecting fake news in social media. 
For example, FANG considers the news outlet that published the article, other news outlets that cited such a publication as citation relationship, and uses different types of stances (i.e., support, deny, comment (neutral) and comment (negative)) that users took when engaging with the news articles as additional information to the engagement graph.

\subsection{Graph Embeddings}
\label{sec:background2}

Recent advances in learning graph structures most commonly use the hyper dimensional space to represent a graph, and represent each node and/or vertex using a vector, based on their neighbouring nodes and the vertices that connect them.
Typically, graph embeddings are achieved by embedding only the topological structure of the graph~\cite{grover2016node2vec,Perozzi2014deepwalk}, or using both the topological structure as well as the auxiliary information of the graph, such as the content of nodes~\cite{chiang2019cluster,hamilton2017inductive}. 

For instance, Node2Vec~\cite{grover2016node2vec} and DeepWalk~\cite{Perozzi2014deepwalk} were the first two neural network approaches developed to represent the graph in a hyper dimensional space. These two methods are similar, as they both use a Skip-Gram architecture with negative sampling to learn the embeddings of each node, based on the portion of graph generated by random walks/edge sampling. 
However, such methods only consider the local context (i.e., closest neighbouring nodes), and cannot obtain a global optimum in representing the complexity of a graph.
To better represent a node in a complete graph structure, graph convolutional networks~\cite{kipf2016semi} (GCN) were proposed to capture the global structure of a given graph. Specifically, GCN uses several layers of graph convolution operations, where each layer is built to capture the information of each node's neighbours,
which is then fed to the next layer, therefore achieving convolutional learning of the graph structure. However, GCN has a very high computational cost~\cite{chiang2019cluster}. Thus, various methods have been proposed to reduce the computational time and space consumption, such as mini-batch stochastic gradient descent with variance reduction GCN~\cite{chen2017stochastic}, and Cluster-GCN~\cite{chiang2019cluster}. 
The GraphSAGE ~\cite{hamilton2017inductive} method was developed to perform parameterised random walks and uses recurrent aggregators. It can be used for both unsupervised or supervised representation learning with a proximity loss between nodes. Moreover, the model adopts a dynamic inductive algorithm, so that it can generate embeddings for unseen nodes and edges at inference time.

Graph embedding methods have been shown to be effective in many tasks, such as node classification~\cite{Perozzi2014deepwalk}, link prediction~\cite{rossi2021knowledge, kazemi2018simple}, and social networks alignment~\cite{du2019joint}.
For example, Monti et al.~\cite{monti2019fake} showed that the user embeddings, trained from a user network with a credibility score for each user, can more accurately identify fake news on Twitter than when the user network information was not used; Wu and Liu~\cite{wu2018tracing} aimed to leverage the community information that users may have encapsulated in their embeddings to identify fake news on Twitter platform. The aforementioned Factual News Graph model (FANG) of Nguyen et al.~\cite{nguyen2020fang} also proposed to use inductive learning for representing social structure, and combined the graphical social network information with sophisticated textual features. Specifically, Nguyen et al. combined citation network among publishers, with user engagement and follower relationship, to build a multi-source graph model for fake news detection on Twitter. 
Similarly, Rath et al.~\cite{rath2020detecting} proposed to use the network structure information to identify the potential super spreader of fake news, and showed the proposed model can identify potential spreaders with the retweet network, the follower-following network, and the timeline data.

\subsection{Summary}
Recent works that aim at detecting fake news most focus on using either textual features and/or static and numerical social network features. 
In this work, we view the users' network structure as a directed graph (i.e., if one user follows/is followed by another user), where users are the nodes in the graph, and their relationships with other users are the edges, thus forms a dynamically incremental graph network.
In order to represent such a directed graph of users, we propose to use graph embedding methods to embed the graph in a hyper-dimensional space, thereby obtaining embedded users representation w.r.t. the graph structure.

Compared to one of the most recent works that adopted graph representation for fake news detection by Nguyen et al.~\cite{nguyen2020fang}, our work also uses inductive representation learning of social network structures in a fake news detection task. However, instead of the complex network information (i.e., the publishers' publication history, citation network among publishers, and the users' followers), we aim to use the \textbf{readily available data} that can be obtained directly through Twitter - without further processing - to construct a user network embedding that is able to accurately cluster users into different groups, based on their friendships with other users and their followers. Building on the users' network embeddings, we then aim to determine if the network structure of users engaging with a news story on social media can effectively represent a news story's truthfulness, and can thus help to detect fake news circulating on social media. Indeed, our later experiments demonstrate that the users' network embeddings do improve the accuracy in detecting fake news in social media. In the next section, we formally define the research problem to be tackled, and present our proposed model, namely the User Network Embedding Structure (UNES) model.

\begin{figure}[tb]
    \centering
    \subfigure[Our proposed UNES model for fake news detection.]{\includegraphics[clip, trim=0.5cm 0cm 0.5cm 0cm,width=0.46\textwidth]{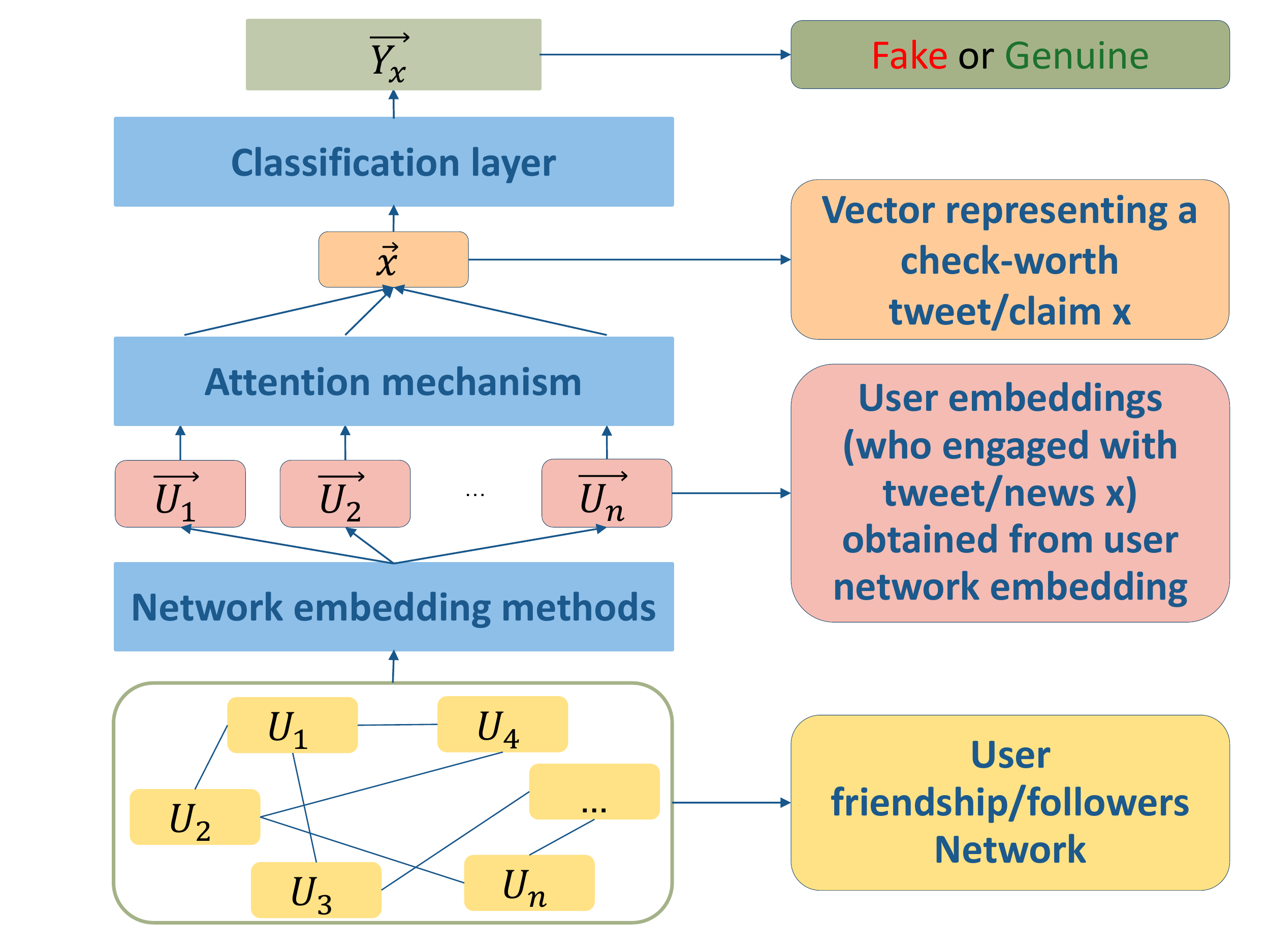}\label{fig:model1}}
    \subfigure[Language model (BERT) baseline for fake news detection.]{\includegraphics[clip, trim=0.5cm 0cm 0.5cm 0cm, width=0.47\textwidth]{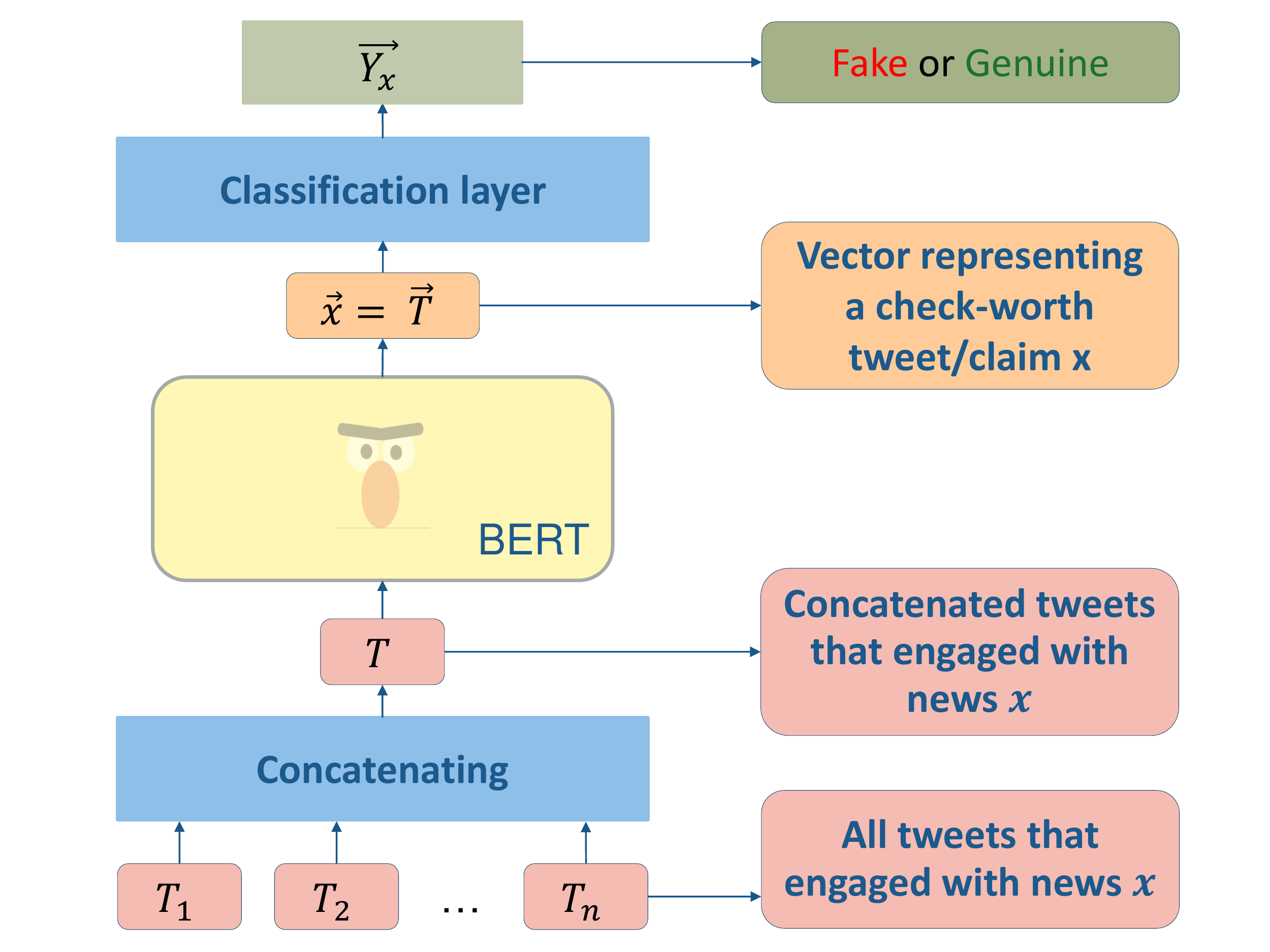}\label{fig:model2}}
    \caption{Comparison between our proposed UNES model and a text-based (BERT) baseline classifier.}
    \label{fig:models} 
\end{figure}

\section{Using Social Network Embedding for Fake News Detection}
\label{sec:method}
In this section, we first define the task we aim to tackle, and describe the notations we will be using in this paper. Thereafter, we introduce our proposed UNES model in detail.

\subsection{Twitter Users-based News Article Classification}

In this paper, we aim to develop a model allowing to accurately classify news articles as fake or not, based on the tweets and retweets with comments that have engaged with the news articles (i.e., included a news link in the tweets), and the corresponding users that posted those tweets on the Twitter SN. In particular, we define the following concepts:

\begin{enumerate}
    \item $X$ = $\{x_1, x_2, ...\}$ is the list of news articles that need to be fact-checked, where each article has a modelled vector representation of each news article $x_i$, denoted by $\overrightarrow{x_i}$.
    \item $U$ = $\{u_1, u_2, ...\}$ is the set of users engaged with all news articles in the dataset. $U_{x_i} \in U$ is the set of users engaged with the specific news article $x_i$. $\overrightarrow{u_n}$ is a modelled vector representation of each user $u_n$.
    \item Specifically, let us define $x = \langle a_{x}, U_{x}, T_{x}\rangle$ where for each news article $x$, $a_{x}$ is the text (i.e., title and body) of the news article, $T_{x}$ is the set of the tweets that engaged with the news article $x$; and $U_{x}$ is the set of users that tweeted $T_{x}$. 
    \item Finally, let $G$ be a representation of the social network graph that is formed by all users $U$.
\end{enumerate}

\noindent Therefore, our news article classification task is defined as follows:
\begin{equation}
\label{eq1}
    \widetilde{Y}_x = f(x) = f(a_x, U_x, T_x)
\end{equation}

In particular, for each news article $x$, we aim to predict whether the article is fake or not, $\widetilde{Y}_x$, based on the text of the article $a_x$,  the tweets $T_x$ that engaged (e.g.\ tweeted/retweeted the news link with comment) with the news article $x$, as well as the users $U_x$ that tweeted $T_x$, where $u_n$ posted the tweet $t_n$ ($t_n \in T_x$). Hence, the objective of this study is to identify the best $f()$ for fake news detection -- next, we describe our proposed UNES model, which identifies fake news on Twitter based on the network structure, $G$, of the Twitter users.

\subsection{Proposed Model - UNES}
\label{sec:model}
We now describe our proposed User Network Embedding Structure (UNES) model for fake news detection on Twitter, which applies user embeddings obtained from network embedding methods. Figure~\ref{fig:models}(a) presents the structure of our proposed UNES model, while Figure~\ref{fig:models}(b) shows a BERT-based model as an example of the state-of-the-art language model for fake news detection.

To obtain a prediction for a given news article $x$ based on the text of the article, the engaged tweets, and the engaged users, i.e.\ $f(a_x, U_x, T_x)$, we propose to make use of the social networks of each of the engaged users $u \in U_x$, as illustrated earlier by Figure~\ref{fig:illu} in Section~\ref{sec:intro}. We consider each user as a node in a directed graph, and their relationships (i.e.\ following, friendship) as vertices connecting with other users. To this end, we introduce $\follows(u_i, u_j)$ as a binary function that returns 1 if user $u_i$ follows $u_j$, and 0 otherwise. Thus user $u_i$ follows user $u_j$ is denoted as $\follows(u_i, u_j)$.
Moreover, Twitter users can be friends -- i.e.\ follow each other -- for which we use two edges that have different directions but connecting the same two nodes:  $\follows(u_i, u_j) \land \follows(u_i, u_j)$.

For all users $U$, the social network graph can be expressed as an adjacency matrix $G$. Specifically, let $G^{\text{fo}}$ denote the followers graph, and $G^{\text{fr}}$ the friends graph. Entries in these adjacency matrices for any pair of users $u_i, u_j \in U$ are defined as $\follows()$:
\begin{equation}
    \begin{split}
        G^{\text{fo}}_{i,j} &= \follows(u_i, u_j)\\
        G^{\text{fr}}_{i,j} &= \follows(u_i, u_j) \land \follows(u_j, u_i)
    \end{split}
\end{equation}
In the following, we use $G$ to represent either $G^{\text{fo}}$ or $G^{\text{fr}}$.

To make use of the large social network graphs within a classifier, we convert the sparse network structure into dense graph embeddings. In particular, the connections of $u_i$ is defined as all users that user $u_i$ is connected to (i.e., as either a follower or a friend). 
Thus, we represent each user $u_i$ by their connections to other users through the application of a graph embedding function\footnote{In the remainder of the paper, we will use graph embedding and network embedding interchangeably, unless it is required otherwise by the context of the section, as we apply graph embedding to the user network structure.} $f_u()$, to obtain embeddings for each user $u$, as follows:
\begin{equation}
    \overrightarrow{u_n} = f_u(G_{u_n})\\
\end{equation}


Note that we employ unsupervised graph embedding models, i.e., users are not labelled with their engagement with fake news or factual news. This is a realistic setup, since a large number of users may not have engaged with any news articles on Twitter, fake or factual.

\begin{table}[tb]
\parbox{.45\linewidth}{
\footnotesize
\centering
\caption{Statistics of the SD dataset. Note that the avg., max., and min. numbers are per news article. \\  \ \\ \ \\ \ \\ }
\begin{tabular}{l|c|c|c}
\hline
                & Overall   & Fake  & Factual \\ \hline
\# of news       & 1054      & 448   & 606\\
\# w/o tweets   & 11        & 4     & 7\\
avg. \# tweets  & 46.30     & 52.31 & 41.89\\
max. \# tweets  & 1054      & 750   & 1054\\
avg. \# users   & 34.10     & 47.48 & 35.95\\
max. \# users   & 1028      & 645   & 1027\\
\hline
\end{tabular}
\label{tab:tweetset}}
\hfill
\parbox{.45\linewidth}{
\footnotesize
    \centering
    \caption{Statistics of the edges users have in our friendships and followers networks, for the SD dataset\protect\footnotemark{}. $\# > 5000$ denotes the users with more than 5000 friends/followers shown in their profile.\\ }
    \begin{tabular}{l|c|c}\hline
    &  \# edges of $G^\text{fr}$ & \# edges of $G^\text{fo}$ \\ \hline
Avg         & 1388.37   &1565.16 \\
Max         & 14012     &18255 \\
Std.        & 1614.43   &1891.05 \\
25\%l       & 181       &147 \\
50\%l       & 674       &621  \\
75\%l       & 2017      &2498 \\
$\# > 5000$ & 3808      &6263  \\ \hline

    \end{tabular}
    \label{tab:con_num}}
\end{table}

We represent each news article as the element-wise sum of all the engaged users' embeddings (i.e., users who tweeted about the news article $x$, as also used in the FANG model~\cite{nguyen2020fang}). Thus, Equation~(\ref{eq1}) can be instantiated as follows:
\begin{equation}
\label{eqf}
    \widetilde{Y}_x = f(x) = f(\sum_{u_n\in U_x} f_x({G_{u_n}}))
\end{equation}

\footnotetext[2]{10 users missing, as 2 accounts are set as private, and 8 accounts were deleted.}
In particular, we use three types of graph embedding methods for the fake news classification task, representing the basic graph embedding models (i.e., DeepWalk), graph cluster-based graph embedding models (i.e., Cluster-GCN), and the existing state-of-the-art graph embedding models (i.e., GraphSAGE), namely:
\begin{enumerate}
    \item \textbf{DeepWalk}~\cite{Perozzi2014deepwalk}. It was the first deep learning-based method to address a graph embedding task, in a manner inspired by word2vec. DeepWalk takes truncated random walks on a graph to learn embedded representations of nodes.

    \item \textbf{Cluster-GCN}~\cite{kipf2016semi}. Given a graph, Cluster-GCN uses a graph convolution operation (GCN)~\cite{kipf2016semi} to obtain node embeddings, by aggregating the neighbouring nodes' embeddings of each node, applying CNN layers for each aggregation. Moreover, Cluster-GCN identifies a subgraph with a graph clustering algorithm, and restricts the neighbourhood search within this subgraph, thus presenting a more efficient and effective graph embedding model than a plain GCN.
    
    \item \textbf{GraphSAGE}~\cite{hamilton2017inductive}. This method performs parameterised random walks and uses recurrent aggregators. It can be used for both unsupervised or supervised representation learning with an algorithm within the model that allows it to generate embeddings for unseen nodes and edges at inference time.
    
\end{enumerate}

Thus, using the graph embeddings from these three methods, we can represent a news article in a hyper-dimensional space based on the social network structure of its engaged users (aggregated as per Equation~\eqref{eqf}), without needing textual information, or a more sophisticated analysis of each users' account (e.g., account type, stance on the topic).
Furthermore, in order to compare the effectiveness of our UNES model with widely-used textual features for detecting fake news on Twitter, in our experiments, we use classifier models learned using social network graph embedding features (as in Figure~\ref{fig:model1}), as well as  those using textual features (as in Figure~\ref{fig:model2}) -- such as the state-of-the-art language model BERT. Next, we detail our research questions and the setup for our experiments.

\section{Experimental setup}
\label{sec:setup}
We aim to address three research questions as follows:
\begin{itemize}
    \item \textbf{RQ1:} Can our proposed UNES model allow to identify clusters of users who engage with fake news on Twitter?
    \item \textbf{RQ2:} How effective is UNES in identifying fake news on Twitter? Which graph embedding method is the most effective?
    \item \textbf{RQ3:} Which type of social network structure (i.e., followers or friendship networks) provides the most effective information in allowing to accurately identify fake news on Twitter?
\end{itemize}

In the following, we describe the used dataset, the approaches we use to represent both tweets and news articles, the user network embedding models, the baselines, and evaluation metrics we use in reporting our results.

\subsection{Dataset}
\label{sec:dataset}
In our conducted experiments, we use the stance detection (SD) dataset provided by~\cite{nguyen2020fang}\footnote{https://github.com/nguyenvanhoang7398/FANG}. The SD dataset consists of news article links and human judgements labels denoting if they are fake or not, as well as engaged tweets, the stance of such tweets, the publisher of the news article, and article citations by other news outlets on Twitter. We download all the available tweets in the dataset, along with the tweet authors' friends and followers. We limit this to a maximum of 5000 for friends and 5000 for followers, which is the maximum number we can download as per Twitter API limit. We also remove those users that have less than 2 edges in the graph (provided they have not engaged with any news article) in order to reduce the size of the graph and allow for tractable experiments.
In Table~\ref{tab:tweetset} we provide the statistics of the dataset in terms of news articles, tweets and users; Table~\ref{tab:con_num} details statistics of the friendship and follower networks of the users.
One can argue that information such as the likes, replies, and retweet relations can also be viewed as possible types of relationships on Twitter. However, due to the difficulties in obtaining data concerning the likes, replies, and retweet relationships from the Twitter API, we do not use these types of relationships in our work. 

\subsection{Semantic Representation}
In order to investigate the effectiveness of social network structure in detecting fake news, we also deploy textual-based classifiers as baselines. In particular, we employ two language processing methods, namely, a TF.IDF representation and a BERT representation for both tweets and news articles. The experimental setup for these two language processing methods are as follows:

\begin{itemize}
    \item \textbf{TF.IDF:} We use NLTK's TweetTokenizer\footnote{https://www.nltk.org/api/nltk.tokenize.html} to tokenise tweets.  Scikit-Learn's  TfidfVectorizer\footnote{https://scikit-learn.org/} is used to extract the TF.IDF representation for both articles and tweets. We limit the maximum number of tokens per text entry to 10k, to include all the tweets and news article tokens. 
    \item \textbf{BERT:} We fine-tune the BERT model with a BERT-base English model (uncased, 12-layer, 768-hidden, 12-heads, 110M parameters). We maintain the suggested learning rate~\cite{devlin2018bert}, with a drop out rate of 0.05, the maximum token length for the news article and tweets combination is 512 (i.e., as the maximum number of tokens that BERT can encode is 512~\cite{devlin2018bert}).
    
\end{itemize}

\subsection{User Network Embedding Methods}
\label{sec:network_emb}

As mentioned earlier, we instantiate \cmn{the} UNES model with three graph embedding methods, namely DeepWalk, Cluster-GCN, and GraphSAGE. Each of these graph embedding methods allows two sets of features per node: node network structures and additional node features. In order to address RQ2, we deploy our models without textual node features. The only information we pass to the graph embedding methods are the network connection features. Moreover, the two types of relationships we use to initiate the graph embedding methods are \textit{friendship} and \textit{followers}. The detailed setup of each graph embedding method is as follows:
\begin{itemize}
    \item \textbf{DeepWalk.} We train DeepWalk with 50 hidden units, a window unit of 10. Each node has a maximum of 10 walks, with a maximum of 80 steps per walk, and results in a 64 dimension vector to represent each user, as per the original DeepWalk paper~\cite{Perozzi2014deepwalk}.
    \item \textbf{Cluster-GCN.} We train Cluster-GCN with 1000 epochs each, with 16 hidden units, and results in 100 dimension vector to represent each user, as per~\cite{nguyen2020fang}.
    \item \textbf{GraphSAGE.} We train GraphSAGE with 30 epochs, 16 hidden units, and 2 layers. It results in 100 dimension vector for each user, as per~\cite{nguyen2020fang}.
\end{itemize}

\noindent \looseness -1 Note that we represent the users who do not have any followers or friends using an embedding vector of [-1, ..., -1].

\subsection{Classifiers}
\label{sec:clf}
We use SVM for the baseline model, along with a TD.IDF representation. We use the Scikit-Learn's implementation of SVM, with the standard parameters (i.e., RBF kernel, a C penalty of 10, and a $\gamma$ of 0.1).
For the deep learning models, we use a fully connected dense layer to classify a news article as fake or factual, which is trained end-to-end with the user embedding vectors. 
For instance, for the BERT model, we fine-tune the BERT model with a fully connected dense layer; for our proposed UNES model, we train the fully connected dense layer as $f_n()$, as per Equation~\eqref{eq1}.

\renewcommand{\labelenumii}{\theenumii}
\renewcommand{\theenumii}{\theenumi.\arabic{enumii}.}

\vspace{-0.5 \baselineskip}
\subsection{Baselines}
\label{sec:baseline}
We report two types of baselines, 5 baselines in total, in order to test the effectiveness of our UNES model.

The first group of baselines consists of models that only use textual features. The second group of baselines are the current state-of-the-art models proposed by~\cite{nguyen2020fang}, namely the GCN model and FANG model. These five baselines are as follows:
\begin{enumerate}
    \item Textual features only:
    \begin{enumerate}
        \item SVM classifier with a TF.IDF representation of tweet and news article, denoted as \textbf{SVM TF.IDF($\boldsymbol{a_n}$ and $\boldsymbol{t_n}$)};
        \item Fine-tuned BERT model, using the news article content $a_n$, denoted as \textbf{BERT($\boldsymbol{a_n}$)};
        \item Fine-tuned BERT model, using the engaged tweets content $T_n$, denoted as \textbf{BERT($\boldsymbol{T_n}$)};
    \end{enumerate}
    \item Models proposed by~\cite{nguyen2020fang}:
    \begin{enumerate}
        \item GCN with FANG network information, as well as TF.IDF representation for the tweets and news, denoted as \textbf{GCN ($\boldsymbol{G^{\text{FANG}}}$)};
        \item GraphSage model using FANG network information, as well as TF.IDF representation for the tweets and news, denoted as \textbf{FANG}. 
    \end{enumerate}
\end{enumerate}

\subsection{Evaluation Metrics}
\label{sec:eval}
We evaluate our methods using the existing training and testing splits (i.e., trained on 10\%, 30\%, 50\%, 70\%, and 90\% of all data) provided by~\cite{nguyen2020fang}. Hence, we test all the models' performances using the same testing sets, and the performances are therefore comparable those numbers reported in \cite{nguyen2020fang}.

As evaluation metrics, we report macro Accuracy, Precision, Recall, and F1, as well as Area under the ROC Curve (AUC) -- indeed, we go further than previous work on the SD dataset, which focused solely upon AUC. 
Moreover, in order to visualise the effectiveness of the users' embeddings in identifying groups of people engaged with fake news versus factual news, we apply the PCA dimension reduction technique on the users' embeddings. For the PCA technique, we also report the cumulative explained variation (CEV), which analyses the variation for individual components, and sums up the variation of the $k$ ($k=50$ in our experiments) most varied principle components. This metric shows the variation between two groups of users (i.e., users that have engaged with fake news vs. users that have never engaged with fake news) that are embedded using the friendship/followers networks.

\begin{figure}[tb]
    \centering
    \subfigure[Friendship network.]{\includegraphics[width= 0.31\textwidth]{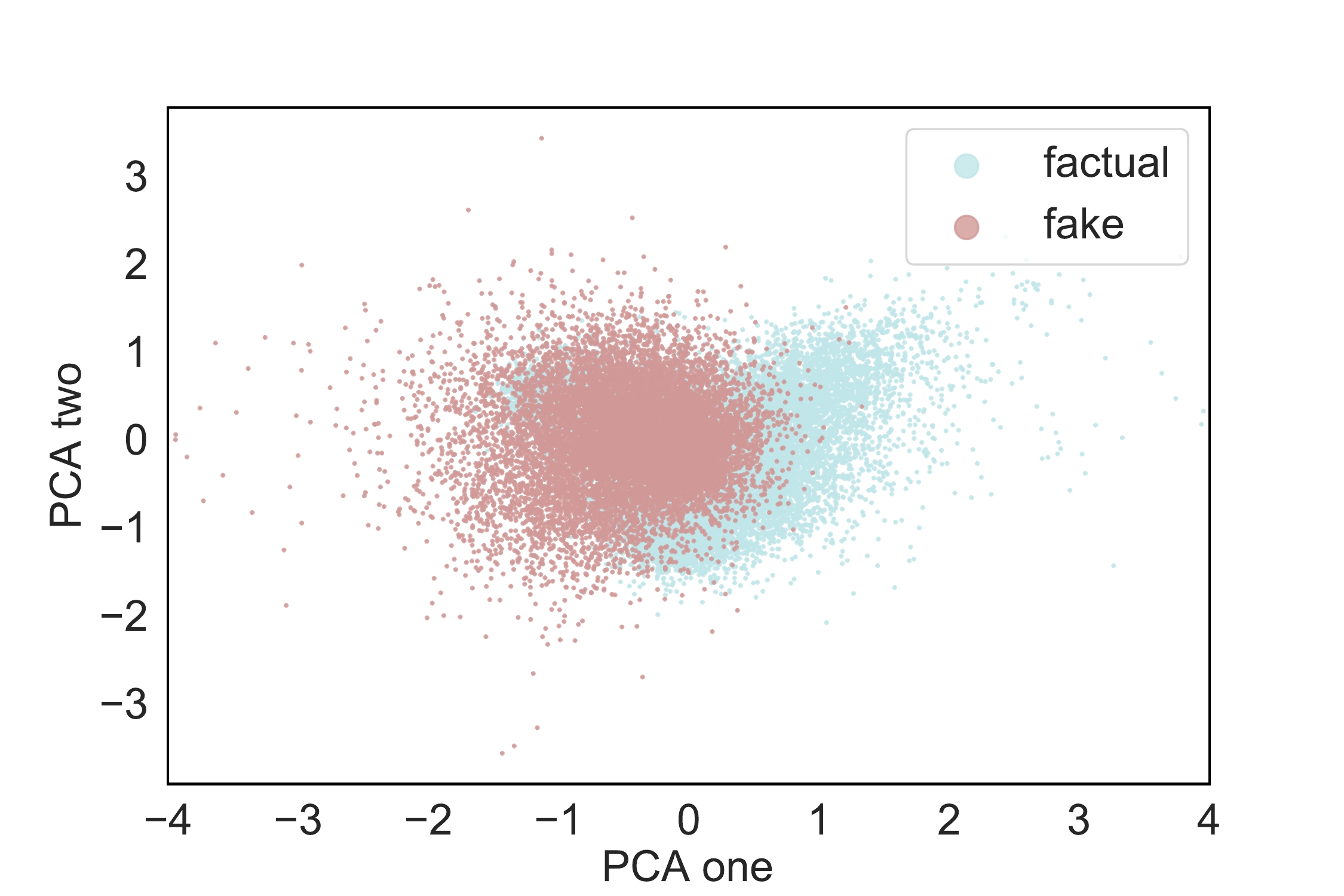}\label{fig:friends_pca}}
    \subfigure[Follower network.]{\includegraphics[width= 0.31\textwidth]{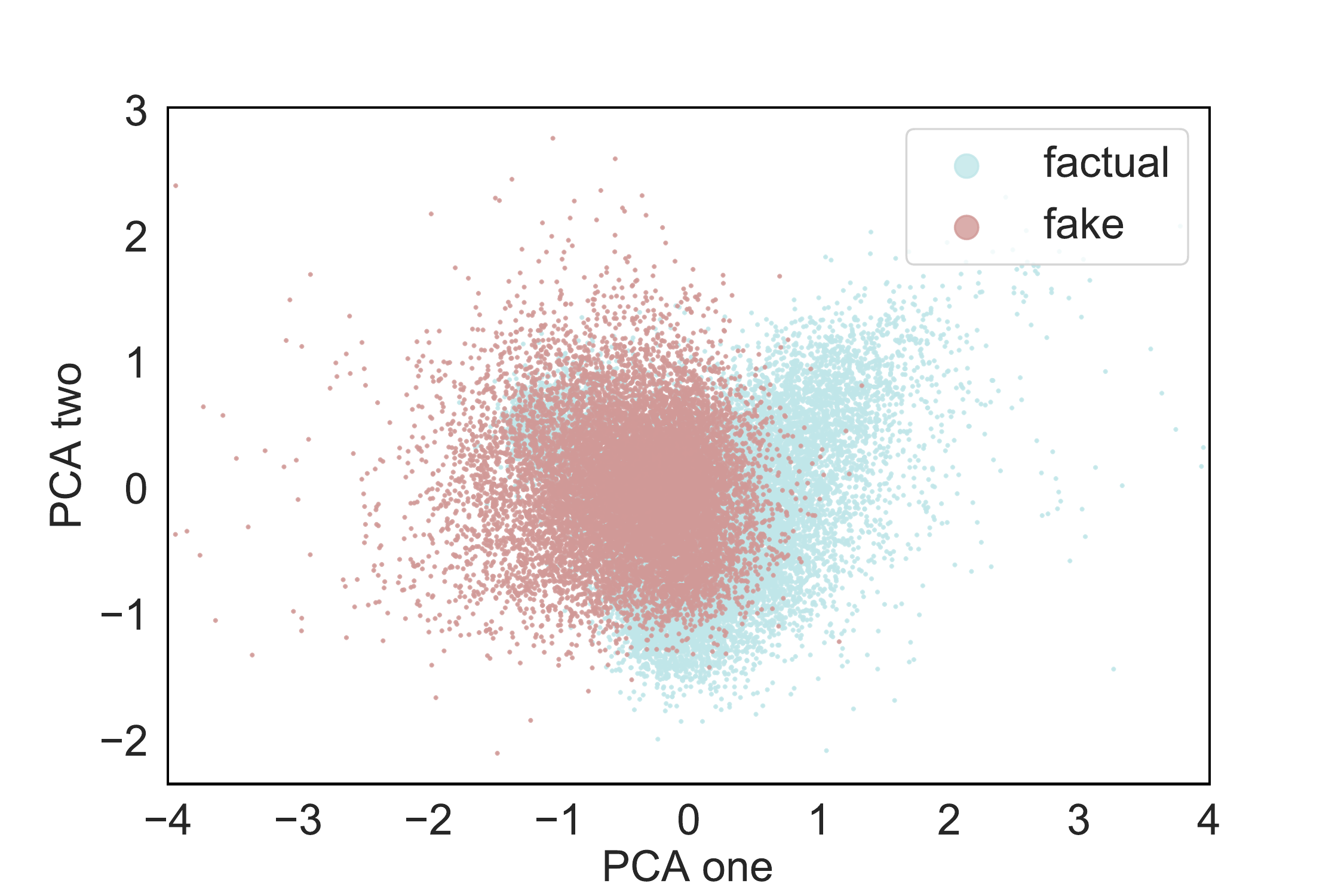}\label{fig:followers_pca}}
    \subfigure[Random user network embeddings.]{\includegraphics[width= 0.31\textwidth]{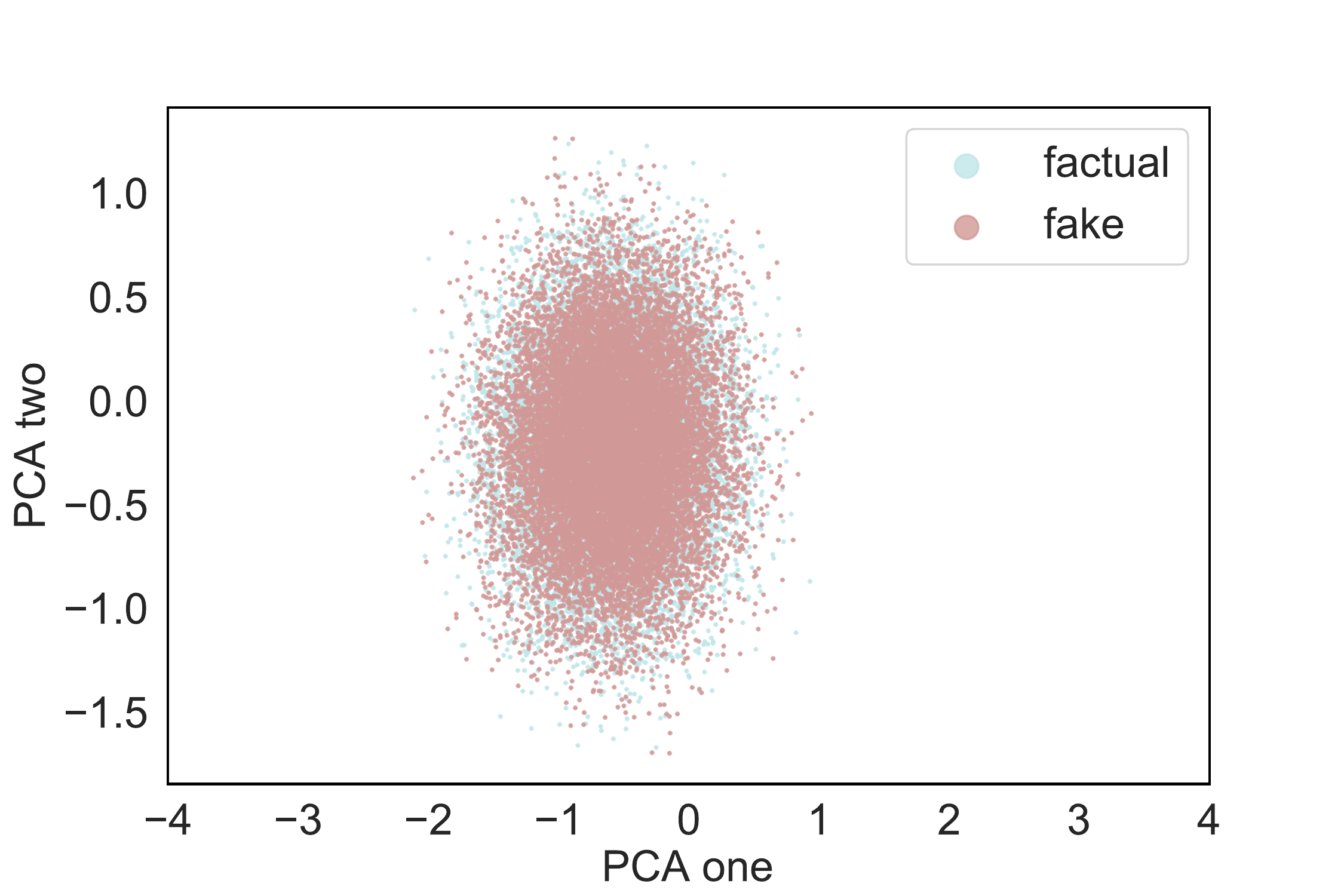}\label{fig:random}}
    
    \caption{Unsupervised embedded users shown in PCA mapping. (a) and (b) are trained using DeepWalk, while (c) uses randomly generated user network embeddings. }
    \label{fig:PCA}  
\end{figure}

\section{Results and Analysis}
\label{sec:results}
In this section, we present the results from our experiments to answer our three research questions. Furthermore, we conduct a case study and discuss the implications of the obtained results.

\subsection{RQ1: Clustering Effect of Users' Network Embeddings}
To gauge the effectiveness of unsupervised user network embeddings in identifying clusters where users engage with fake news, we visualise the distributions of users in our embedded user networks, using the PCA dimensionality reduction technique\footnote{Plots using the tSNE dimensionality reduction technique produced similar observations, and hence are omitted for reasons of brevity.}, and measure the cumulative explained variation (CEV) between users who engage with fake news and users who engage with factual news. Specifically, we visualise the users' embeddings trained with the DeepWalk graph embedding method, since DeepWalk provides a lower bound in user embeddings performance, due to its simplicity compared to the Cluster-GCN and the GraphSAGE methods.

Figures~\ref{fig:friends_pca} \&~\ref{fig:followers_pca} illustrate the distribution of user embeddings (learned from the friendship network and the followers network with the DeepWalk graph embedding method) that are reduced to 2 dimensions using a PCA. 
The red dots denote users that have engaged with at least one fake news article, and the teal dots denote users that \textbf{only} engaged with factual news articles. 
Note that as mentioned earlier, during the training session, we did not label users as engaged with factual news or engaged with fake news, hence the embeddings are learned in an unsupervised fashion. In Figure~\ref{fig:random}, we also show randomly generated user embeddings plotted using PCA, where, unlike in Figures 3a and 3b, the distributions of users engaging with fake news and with factual news are more uniform.

Specifically, from Figure 3a, it can be observed that for the user embeddings learned from the friendship network, users who are engaged with fake news are more tightly clustered together.
That is, the PCA mapping shows that users who engaged with fake news are tightly gathered around the top left corner, which suggests that the users who engage in fake news are more tightly grouped into smaller echo chambers than the users who do not engage in fake news, as observed by Yoo~\cite{yoo2007ideological}. 
Moreover, the CEV analysis shows that the top 50 principle components achieved a cumulative explained variation of 0.8472, indicates that the two groups of users have a variance of 84.74\% calculated from the embeddings, learned from the friendship network.
In Figure~\ref{fig:followers_pca}, we observe similar trends from the user embeddings learned from the followers network, with the CEV for the 50 principle components being 0.8472, indicates that there are valid differences between the two user groups, in comparison to the randomly generated user embeddings, where the CEV is 0.5863.
Our results echo the findings in~\cite{tornberg2018echo}, namely that users tend to form a more tightly grouped community when they are more heavily influenced by fake news\final{, and \bof{show}} that echo chamber effects indeed exist in social media. 

Thus, in response to RQ1, we conclude that the unsupervised Twitter user network embeddings can indeed cluster users into different groups (i.e., users who have engaged with fake news, versus users who only engaged with factual news), with a cumulative explained variation of around 85\%, using either the follower network or the friendship network.

\begin{table*}[tb]
\footnotesize
\centering
\caption{Performance comparison among the models using 90\% training data. Numbers in the significance column indicates that the model is significantly better than the numbered model (McNemar's Test, $p$$<$$0.01$). }
\label{tab:r90}
\begin{tabular}{c|l|l|l|l|l|l|l} 
\hline
\# & Model                          & Accuracy & P      & R      & F1     & AUC         & Significance  \\ \hline
\multicolumn{8}{c}{\new{Textual }Baselines -- Textual features only}                                                   \\  \hline
1  & Random                         & 0.4737   & 0.3205 & 0.5000 & 0.3906 & 0.5000      & -             \\
2  & SVM TF.IDF($a_n$ and $T_n$)    & 0.6068   & 0.6010 & 0.6095 & 0.5962 & 0.6095      & 1, 4             \\
3  & BERT($a_n$)                    & 0.5897   & 0.5584 & 0.5595 & 0.5588 & 0.5595      & 1, 4             \\
4  & BERT($T_n$)                    & 0.5299   & 0.5410 & 0.5443 & 0.5249 & 0.5443      & 1             \\ \hline
\multicolumn{8}{c}{Network baselines from FANG -- \new{Complex network features and texutal features}} \\
\multicolumn{8}{c}{\new{(publisher, citation, follower network, stance and TF.IDF features)}}            \\ \hline
5  & GCN($G^{\text{FANG}}$)              & 0.6458   & 0.6328 & 0.6250 & 0.6262 & 0.7125      & 1-4, 7   \\ 
6  & FANG                           & 0.6875   & 0.6799 & 0.6821 & 0.6807 & \textbf{0.7518}    &  1-5, 7,9   \\  \hline
\multicolumn{8}{c}{UNES variants -- \new{Unsupervised network features only}}                   \\  \hline
7  & DeepWalk($G^{\text{fr}}$)             & 0.6410   & 0.5717 & 0.5052 & 0.4122 & 0.5052      & 1-4      \\
8  & Cluster-GCN($G^{\text{fr}}$)          & 0.7083   & 0.7083 & 0.7142 & 0.7062 & 0.7071      & 1-7,9    \\
9  & Cluster-GCN($G^{\text{fo}}$)          & 0.6667   & 0.6556 & 0.6500 & 0.6515 & 0.7000      & 1-5,7    \\
10 & GraphSAGE($G^{\text{fr}}$)            & \textbf{0.7708}   & \textbf{0.7650} & \textbf{0.7607} & \textbf{0.7625} & 0.7498      & 1-9,11  \\
11 & GraphSAGE($G^{\text{fo}}$)            & 0.7292   & 0.7222 & 0.7250 & 0.7233 & 0.7365      & 1-9      \\
\hline
\end{tabular}
\end{table*}

\begin{figure}[tb]
    \centering
    \subfigure[Accuracy.]{\includegraphics[width=0.31\textwidth]{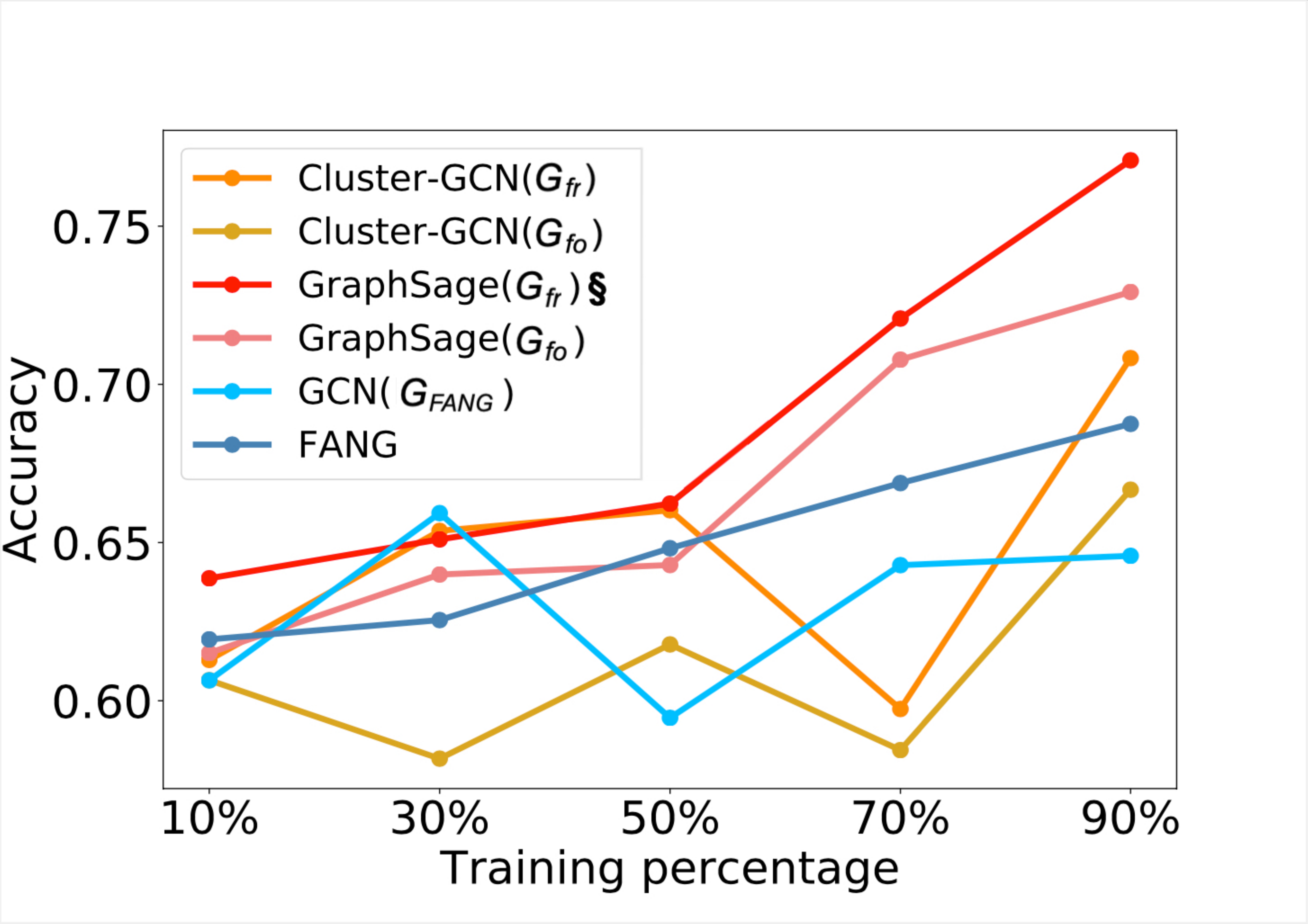}\label{fig:acc}} 
    \subfigure[F1 score.]{\includegraphics[width=0.31\textwidth]{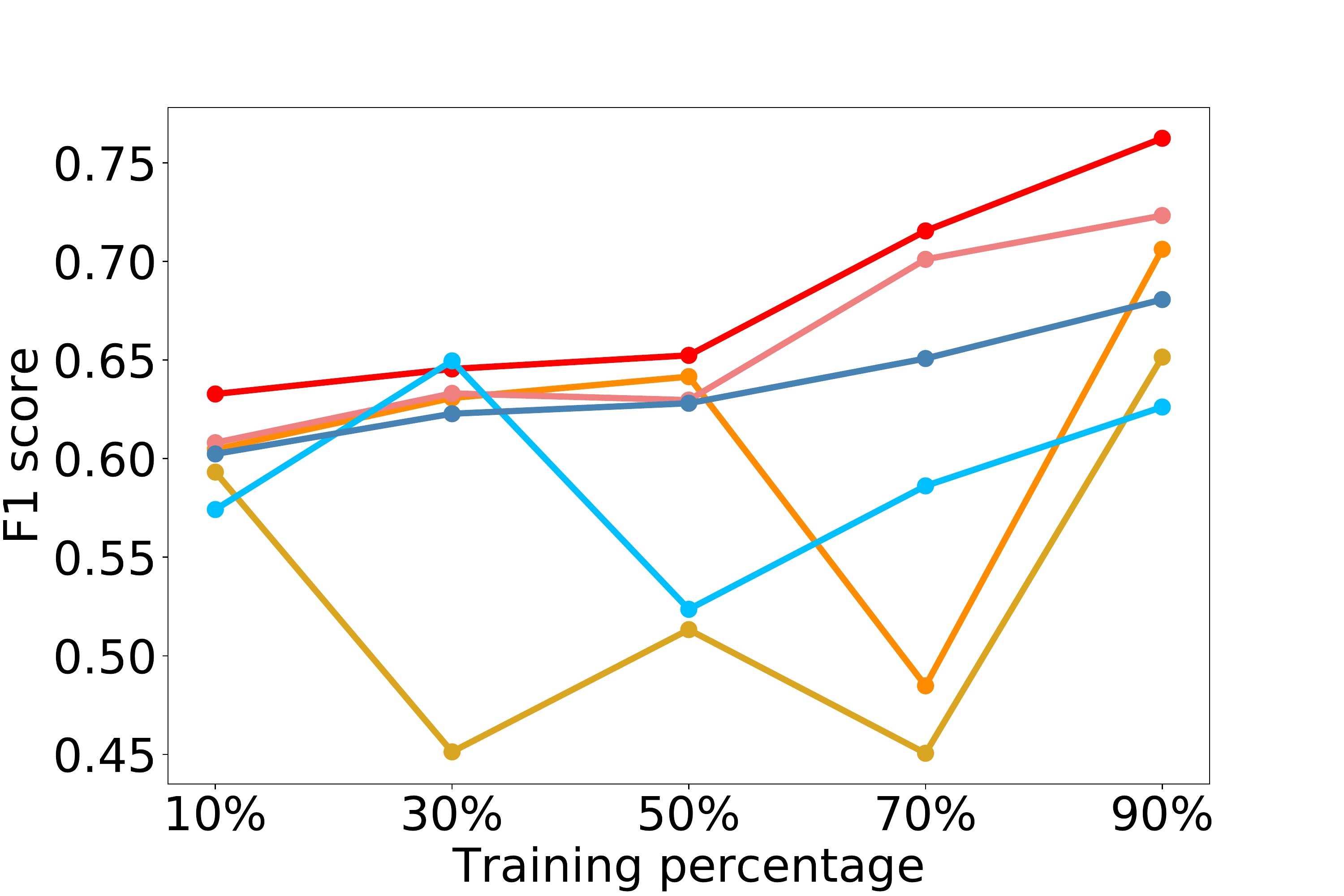}\label{fig:F1}}         
    \subfigure[ROC AUC]{\includegraphics[width=0.31\textwidth]{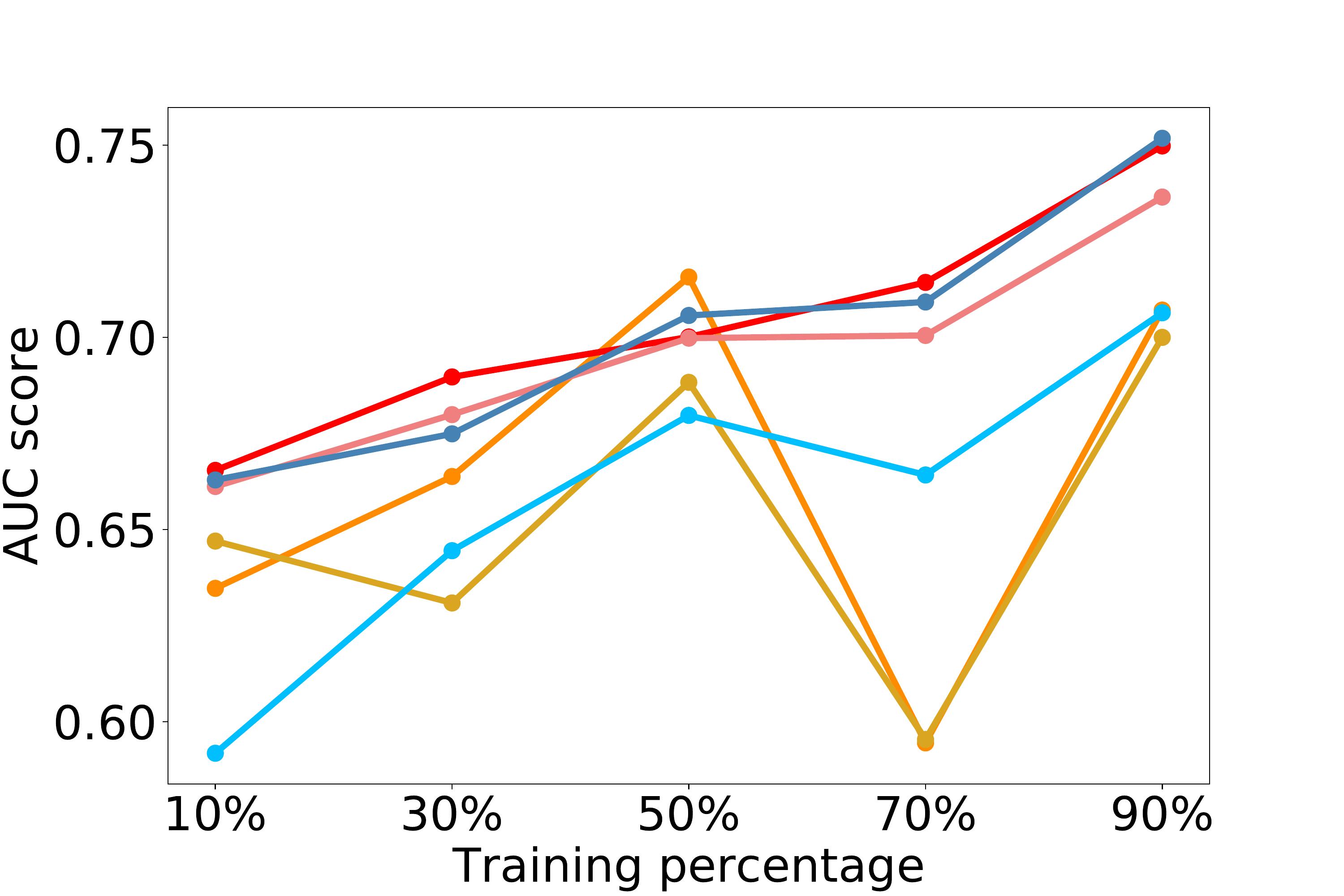}\label{fig:auc}} 
    \caption{Performances of GraphSAGE($G^{\text{fr}}$), GraphSAGE($G^{\text{fo}}$), 
    Cluster-GCN($G^{\text{fr}}$), Cluster-GCN($G^{\text{fo}}$), GCN($G^{\text{FANG}}$), and the FANG model, on accuracy, F1, and AUC. 
    \S \ in the legend denotes that our corresponding model variant significantly outperforms FANG, on all training percentages.}
    \label{fig:split} 
\end{figure}

\subsection{RQ2: UNES Model for Fake News Classification}

To evaluate whether our proposed UNES model is more effective in detecting fake news on the SD dataset w.r.t. the baseline models, as well as to identify the most effective graph embedding method for fake news detection on Twitter, we compare our various UNES instantiations with the baselines described earlier in terms of classification performances.
Table~\ref{tab:r90} shows how the models perform when trained with 90\% of all data, validated with 5\% of the data, and tested on 5\% of the data. Indeed, we use the pre-partitioned training-testing sets provided by \cite{nguyen2020fang}, thus the results shown in Table~\ref{tab:r90} are comparable to those reported in~\cite{nguyen2020fang}. To analyse the effect of the size of training data on the results, we also evaluate the various models across different training data size percentages, and present the results in Figure~\ref{fig:split}, in terms of Accuracy, F1 and ROC AUC.

On analysing Table~\ref{tab:r90}, we firstly observe that the text-based baselines (i.e., the TF.IDF and BERT models, lines 2-4 in Table~\ref{tab:r90}) can identify fake news significantly better the random baseline. 
However, a simple SVM(TF.IDF) model outperforms both BERT models, indicating the difficulties for contextual features to analyse news and tweets on Twitter, with the SD dataset. 
Moreover, all models that use network features (lines 5-11) significantly outperform the textual features only models (lines 1-4). This observation suggests that network features are more successful at identifying fake news on Twitter than textual features alone.

Secondly, we observe that with only the network information, our UNES model with the simpler DeepWalk graph embedding (line 7 in Table~\ref{tab:r90}) does not outperform either of the network baselines (lines 5 \& 6).
On the other hand, contrary to DeepWalk, the use of the UNES model along the advanced graph embedding models (i.e., Cluster-GCN and GraphSAGE) can achieve better accuracy and F1 performances, compared to the GCN($G^{\text{FANG}}$) and FANG, respectively.
Indeed, Cluster-GCN with friendship/followers networks (lines 8 \& 9) can significantly outperform GCN($G^{\text{FANG}}$) (line 5), while GraphSAGE with the friendship/followers networks (lines 10 \& 11) significantly outperforms FANG (line 6).
However, FANG (line 6) obtained the highest ROC AUC performance among all the tested models. A further inspection of all the evaluated UNES variants (i.e., lines 7-11 in Table~\ref{tab:r90}) shows that while they are accurate, they are generally less certain in their predictions. Specifically, the dense neural network layer outputs posterior probabilities for each class closer to 0.5 rather than 0 or 1 for our binary classification task (i.e., classifying a given news as fake or not), while the FANG baseline tends to produce probability outputs closer to 0 and 1 rather than 0.5. We argue that this is because the UNES model uses a network structure data with no textual information on the content of the news or tweets, unlike FANG, which uses both the textual data (the stance and TF.IDF representations of the news articles and the tweets) along with the network data. We leave to future work, the investigation of how best to integrate the textual data into the UNES model. 

Furthermore, Figure~\ref{fig:split} shows the performances of the UNES variants and the FANG baselines when tested with all possible training data size percentages. 
From the figure, we observe that the UNES variant using the GraphSAGE($G^{\text{fr}}$) graph embeddings consistently outperforms FANG in terms of the accuracy and F1 metrics, regardless of the used training data size. 
However, similar to the observation that FANG achieved the highest ROC scores in Table~\ref{tab:r90}, we observe that all the variants of our UNES model (i.e., Cluster-GCN($G^{\text{fr}}$), Cluster-GCN($G^{\text{fo}}$), GraphSAGE($G^{\text{fr}}$), and GraphSAGE($G^{\text{fo}}$)) do not outperform FANG consistently, due to the aforementioned issue, namely that model GraphSAGE($G^{\text{fr}}$) tends to predict posterior probabilities closer to 0.5 rather than 0 or 1.

Moreover, from Figure~\ref{fig:split} we observe that all the GCN-based models (i.e., Cluster-GCN($G^{\text{fr}}$), Cluster-GCN($G^{\text{fo}}$), and GCN($G^{\text{FANG}}$)) suffer from instability when trained using different percentages of training data. Specifically, the Cluster-GCN($G^{\text{fr}}$) and Cluster-GCN($G^{\text{fo}}$) approaches both exhibit performance drops when trained with 30\% and 70\% of the data. One of the baseline models, GCN($G^{\text{FANG}}$), shows a drop in the Accuracy and F1 performances when trained with 50\% of the data, and a ROC AUC drop when trained with 70\% of the data.
Indeed, the GCN-based models have unstable performances when tested on the different training percentages. This suggests that the GCN graph embedding model might represent nodes (i.e., users in our experiments) inaccurately as embeddings, as GCN and Cluster-GCN both identify subgraphs before computing the embeddings of each node, while the subgraphs are not updated throughout the training session, contrary to the GraphSAGE model,  which aims to perform graph embeddings without any subgraph partitioning. 



Overall, in response to RQ2, we conclude that the models that use the users' network embeddings alone significantly outperform the language model baselines in classifying fake news on the SD dataset. Moreover, our models that use unsupervised users' network embeddings can identify fake news on the SD dataset more accurately than the FANG baseline, which uses complex users' network embeddings that include additional relations such as the publisher network and the citation network, as well as the textual information from the tweets and news articles. Among the variants of our proposed UNES model, the variant using the GraphSAGE graph embeddings is the most effective.

\subsection{RQ3: Followers or Friends?}
Recall that users can have two types of relations with other users, namely through the following relation or through the friendship relation, where the latter requires both users to follow each other. As discussed before, we instantiate the UNES model using either the follower or the friendship relationships for both the GraphSAGE and Cluster-GCN graph embedding models. Their respective performances are included in Table~\ref{tab:r90} and Figure~\ref{fig:split}. On analysing Table~\ref{tab:r90}, we observe that with 90\% of the data as training data, GraphSAGE($G^{\text{fr}}$) outperforms GraphSAGE($G^{\text{fo}}$) on all metrics, while Cluster-GCN($G^{\text{fr}}$) outperforms Cluster-GCN($G^{\text{fo}}$) on all metrics. Figure~\ref{fig:split} shows that GraphSAGE($G^{\text{fr}}$) outperforms GraphSAGE($G^{\text{fo}}$) consistently on all metrics, regardless of the portion of training data used. 

To understand this result, we analyse the statistical differences between the friendship network and the follower network in the SD dataset to investigate the dissimilarity of their embeddings. Recall Table~\ref{tab:con_num}, where it can be seen that on average, users have a higher number of edges in the follower network than in the friendship network (1565.16 vs. 1388.37), showing that on average users have more followers than friends, echoing the fact that the friendship network on Twitter forms a more sparse network than the follower network. 
Although, intuitively speaking, the denser the network, the more information we can collect, we argue that the denser followers network may introduce more noise due to the lack of shared interests between followers and followees, compared to friends that share more interests and similar opinions, as represented by the friendship network.

Furthermore, the higher accuracy of the friends graph is advantageous from a data point of view: as noted that we are limited in terms of possible Twitter API calls, which reduces the observable portion of the users' friends or follower networks. 
Indeed, from Table~\ref{tab:con_num} it can be observed that the friendship network has a smaller proportion of users with more than 5000 friends (i.e., 10.64\% of all news-engaging users in the SD dataset do not have all their friends downloaded); in contrast, the follower network has 6263 users with more than 5000 followers (17.50\% of all news-engaging users). 
Overall, in response to RQ3, we conclude that in our experiments, with a limited number of Twitter API calls available, the users' friendship network is the most effective type of social network relationships, when used to construct the users' graph embeddings and for detecting fake news circulating on Twitter.

\begin{table}[tb]
\footnotesize
\centering
\caption{Case study examples. The \textit{Fake?} column indicates if the news article is fake news (Y) or not (N); \# $T_n$ describes the number of tweets are associated with the news article.}
\begin{tabular}{c|c|l|c}\hline
Case & Fake? & News article title & \# $T_n$\\ \hline 
\multicolumn{3}{c}{UNES correct, BERT incorrect} \\\hline 
1 & N & At Trump hotel site, immigrant workers wary & 462 \\
2 & N & Charlie Hebdo : Le témoignage de la dessinatrice Coco. &132 \\
3 & N & Coldsip.com | News & 155 \\\hline 
4 & Y & Illegal Immigrant Deported 6 Times Charged in Felony Hit-and-Run of Family that Injured Children. & 588\\
5 & Y & Sasha Obama Just Crashed Her Expensive New Car Into A Lake. & 246 \\ 
6 & Y & Hundreds of people died after eating the Patti LaBelle brand Patti Sweet Potato Pie. & 227 \\ \hline
\multicolumn{3}{c}{UNES incorrect, BERT correct}  \\\hline
7 & N & UCLA Students protest after partygoers wear blackface at fraternity party. &246 \\
8 & N & Killer to face firing squad.& 539 \\
9 & N & Caitlyn Jenner to receive courage award at ESPY’s & 797\\ \hline
10 & Y & Obama Orders Chicago School to Let `Transgender' Boy Use Girls' Locker Room. & 398\\

11 & Y & Peanut Butter and Jelly Deemed Racist. & 835\\
12 & Y & UPDATE: Second Roy Moore Accuser Works For Michelle Obama Right NOW. & 3913 \\ \hline
\end{tabular}
\label{tab:case} \vspace{-\baselineskip}
\end{table}

\begin{figure}[tb!]
    \centering
    \includegraphics[width=0.5\textwidth]{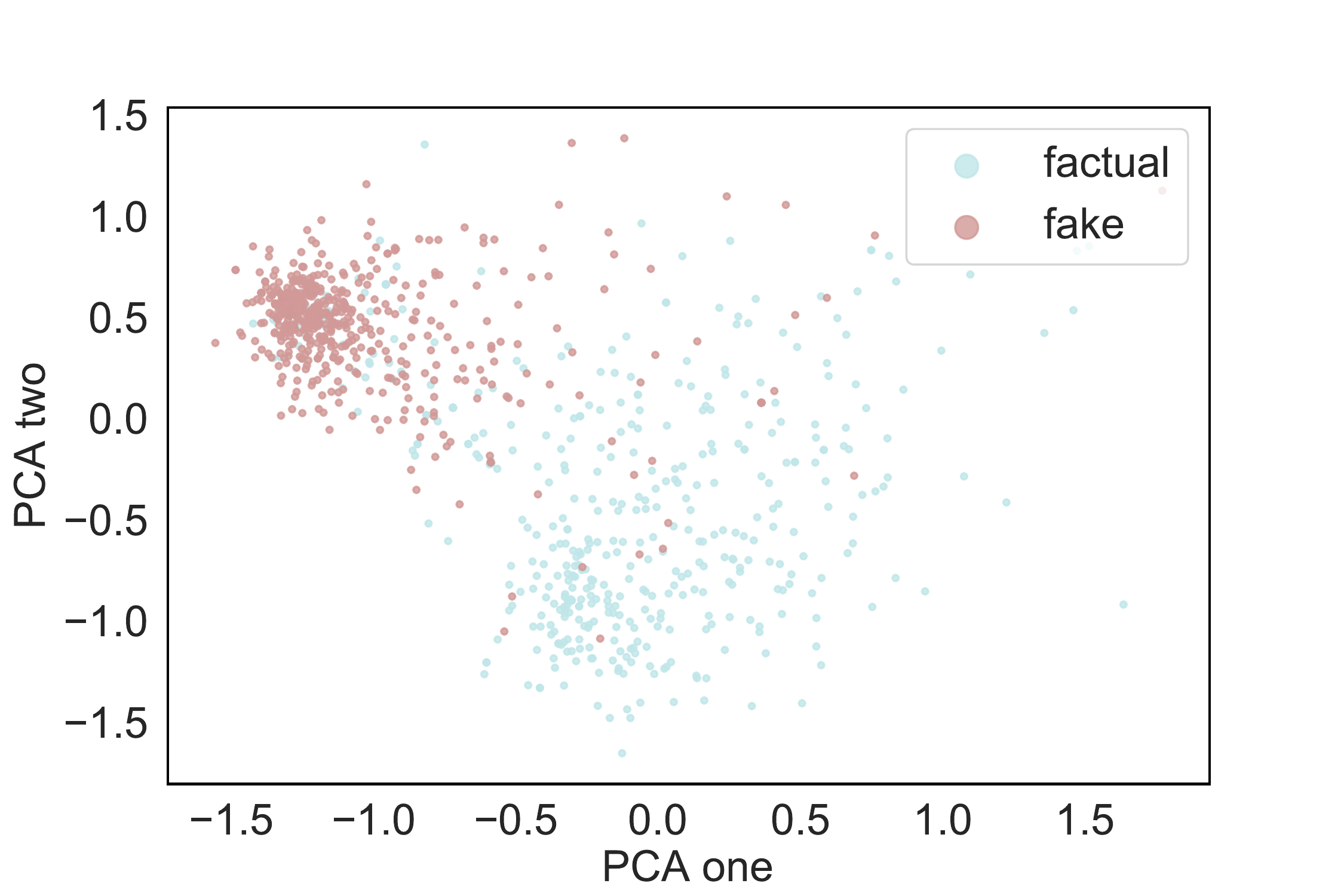}
     \captionsetup{width=0.85\linewidth}

    \caption{PCA mapping of the users engaged with 2 news articles related to immigration issues, on Friendship network. Red dots represent users who engaged with fake news ``\textit{Illegal Immigrant Deported 6 Times Charged in Felony Hit-and-Run of Family that Injured Children}'', while teal dots represent users that engaged in factual news ``\textit{At Trump hotel site, immigrant workers wary}".}
    \label{fig:pca_case}
\end{figure}

\subsection{Case Study}
In order to better demonstrate the successes and failures of our proposed UNES model, we present in Table~\ref{tab:case} a case study. In the table, in cases 1-6 UNES makes a correct prediction where the BERT model does not, while in cases 7-12 are for the opposite condition. We also report the number of tweets discussing each news article.

\looseness -1 Indeed, from cases 1-3 in the table, we observe that our UNES model can classify the news correctly regardless of the language (case 2), or if the title is redacted (case 3). These cases show that our proposed model can be used universally across different language, and when the news content is corrupted. Cases 4-6 in Table~\ref{tab:case} shows that the language used in the news title can mislead the language models into misclassifying the news as genuine, while our proposed UNES model can correctly classify the news as fake with additional information from the users engaged in such news. 

On the other hand,  cases 7-12 show difficult cases for the UNES model. These cases are mostly related to significant and controversial topics (such as racism, politician scandals, transgender right, immigrants), where people with different views can be easily drawn to such news, and voice their opinions. We believe these cases are indeed difficult to identify with a cluster-based model. 

Furthermore, case 1 (factual) and case 4 (fake) are two news articles related toimmigration issues in the US, that the UNES model correctly classified. Figure~\ref{fig:pca_case} shows the PCA mapping of the users engaged in these two cases. We observe two embedding clusters from Figure~\ref{fig:pca_case}, one among users who commented on the fake news (case 4) and one among users who commented on the factual news (case 1). These two user clusters illustrate a clearer echo chamber effect on the immigration issue, than the lack of separation among the clusters observed in Figure~\ref{fig:friends_pca}. We believe these two clearly separated user clusters explain why UNES accurately classified these two cases. In order to further improve accuracy of identifying fake news, in the future we aim to combine UNES with language models to enhance the classification accuracy in such difficult examples. 

\section{Conclusions and Limitations}
\label{sec:conclusions}
In this paper, we have \textbf{\textit{demonstrated}} that:
\begin{itemize}
    \item User network embedding methods can cluster the users who engaged in fake news more closely than users who only engaged with factual news, observing the echo chamber effects, with $CEV \approx 0.84$;
    \item Our proposed UNES model, using unsupervised network embedding model trained with the users' relationships, can identify the users clusters that engage with fake news, when tested on a large recent Twitter dataset; 
    \item With the readily available information (i.e., only the followers/friendship networks), the UNES model can identify fake news more accurately than the existing SOTA models such as FANG (uses complex network information that requires additional labelling) and text-based models; 
    \item The friendship network can more accurately help identify fake news than the follower network, within the limited access afforded to the Twitter social networks given the available Twitter API rate. 
\end{itemize}

\noindent However, we also identify several \textbf{\textit{limitations}}:
\begin{itemize}
    \item The current UNES model does not perform well when combined with language representations, thus it’s important to identify better mechanisms that can leverage both types of information (i.e., user network embeddings and language embeddings);

    \item User networks are dynamic and are constantly changing, how to adapt to the changing user network representation is worth studying, thus help identifying new types of fake news clusters and their associated fake news;
    \item Develop methods to better handle cold-start users (i.e., the new accounts without social connections and/or tweets). 
\end{itemize}

\bibliographystyle{unsrt}  
\bibliography{templateArxiv}

\end{document}